\documentclass[aps,pra]{revtex4}

\usepackage{amsmath}
\usepackage{epsfig}
\usepackage{array}
\usepackage{times}
\usepackage{setspace}
\usepackage{mathrsfs}
\usepackage{amssymb}
\usepackage{color}
     
\begin{document}

\title{Work extraction from heat-powered quantized optomechanical setups}

\author{D. Gelbwaser-Klimovsky}
\altaffiliation{Present address: Department of Chemistry and Chemical Biology, Harvard University, Cambridge, MA 02138.}
\affiliation{Weizmann Institute of Science, 76100
Rehovot, Israel}
\email[Correspondence to:]{dgelbi@yahoo.com.mx}
\author{G. Kurizki}
\affiliation{Weizmann Institute of Science, 76100
Rehovot, Israel \\
}

\begin{abstract}
We analyze  work extraction from an autonomous (self-contained) heat-powered optomechanical  setup. The initial state of the quantized mechanical oscillator plays a key role. As the initial mean amplitude of the oscillator decreases, the resulting efficiency increases. In contrast to laser-powered self-induced oscillations, work extraction from a  broadband heat bath does not require coherence or phase-locking: an initial phase-averaged coherent state of the oscillator still yields work, as opposed to an initial Fock-state.
\end{abstract}
\maketitle

Quantum optomechanics has exhibited tremendous theoretical and experimental progress in recent years towards controlled manipulations of the interaction of cavity photons with mechanical oscillators  \cite{manciniprl98,manciniprl02,vitaliprl07,aspelemeyerarxiv13,meystreann13,chenjpb13,teufelnat11,ChanNAT11,brooksnat12,safaviarx13,purdysci13,palomakinat13,lorchprx14,schliessernatphys08,EpsteinNAT95,kippenbergprl05,carmonprl05,metzgerprl08s,anetsbergernat09,zaitsevpre11,marquardtprl06,zaitsevnondyn12,khurginprl12,khurginnjp12,ludwignjp08,vahalapra08,rodriguesprl10,armourcrp12,qianprl12,nationpra13,genespra08}. A prominent application of such manipulations has been the cooling of the mechanical oscillator \cite{teufelnat11,ChanNAT11,schliessernatphys08} and the converse regime of the  amplification of its motion \cite{kippenbergprl05,zaitsevpre11,marquardtprl06,zaitsevnondyn12,khurginprl12,khurginnjp12}. Provided  the total damping rate of the oscillator is negative, the laser-driven cavity mode can parametrically amplify the mechanical motion in a self-sustained limit-cycle of oscillations\cite{kippenbergprl05,carmonprl05,metzgerprl08s,anetsbergernat09,grudininprl10,zaitsevpre11,marquardtprl06,zaitsevnondyn12,khurginprl12,khurginnjp12} that has a nonclassical counterpart \cite{ludwignjp08,vahalapra08,rodriguesprl10,armourcrp12,qianprl12,nationpra13}.

Here we explore a different avenue, by raising  the question: can the quantized oscillator transform thermal energy from a heat bath (rather than a laser) into mechanical work \cite{CallenBOOK85,GemmerBOOK10,Kosloffentr13}  and thus \textit{act as a heat engine?} Is such work similar to the self-induced oscillation discussed above? These questions pertain to a subtle and unsettled issue: what is the \textit{proper definition of work} in a quantized (non-driven) heat engine and what limitations does  thermodynamics set on its extraction?

We address these issues in the context of  a realizable model that  consists of a working medium, here the optical mode, constantly  coupled to two distinct  thermal baths and the  mechanical oscillator that extracts the work.  This is the standard situation in quantum open systems: our cavity mode  constantly  and unavoidably  interacts with the outside electromagnetic vacuum (the cold bath) and with a spectrally filtered heat source (hot bath). This setup is  an example of a \textit{continuous} heat machine  that \cite{Kosloffentr13}, contrary to more commonly known strokes-operated machines (such as the Carnot or Otto engine) \cite{aspelemeyerarxiv13,AlickiJPA79,GemmerBOOK10,Kosloffentr13,PalaoPRE01,LindenPRL10,LevyPRL12,BrunnerPRE12,correaPRE13,zhangPRL14}, does not involve decoupling from and recoupling to alternate (hot and cold) thermal baths. Continuous, fully quantized machines may therefore be operated completely autonomously, without external intervention, after launching them in a ``push $\&$ go'' fashion. 
While there is a well-established definition of work as long as the heat machine   is driven by an external field (acting as a piston) \cite{Kosloffentr13,AlickiARXIV12,AlickiJPA79}, it is more subtle to quantify work once this field is quantized. This subtlety is related to   the fundamental question:  How is the energy exchange between  two quantum systems  divided between heat and work?

 Here we  invoke the general and rigorous definition of maximal work storage (capacity) in the device: it is  measured by the nonpassivity (see below) \cite{LenJSP78,puszcmp1978,gelbiepl13} of the quantum state of the ``piston'', here the mechanical oscillator. The "piston" interacts with the  working medium (here the optical mode) , which in turn is thermalized by the two baths.  As the piston evolves, typically in a non-unitary fashion, its maximal work capacity (nonpassivity) changes. This rate of change is the maximal extractable power.
 In driven-piston scenarios  \cite{AlickiJPA79,GemmerBOOK10,Kosloffentr13,PalaoPRE01,LindenPRL10,LevyPRL12,BrunnerPRE12,correaPRE13,zhangPRL14,Skrzypczykarxive13,procacciajcp76,AlickiARXIV12}  work can be extracted  by the piston from the system via unitary or classical operations. It is then independent of the initial state of the piston, which is not a thermodynamic resource. By contrast, we are concerned with work capacity in an autonomous, quantized setup that crucially depends on the initial state of the oscillator – just as the work stored in an initially compressed spring. This initial state is then an extra thermodynamic resource quantified by nonpassivity, which is largely ( but not completely) determined by its negentropy (see SI). This extra resource may be crudely viewed as an additional (pseudo) "bath" whose state-dependent effective  temperature $T_M$  may be ( for some time) arbitrarily low.  Consequently, this extra resource may yield higher efficiency than the standard Carnot-cycle efficiency 1-Tc/Th, where the only resources are the hot and cold baths at temperatures Th and Tc, respectively. Consistency with the Second Law is ensured by construction ( see SI) .
Not less  important is the validity of  this analysis for \textit{arbitrary nonunitary  and  nonadiabatic evolution }\cite{AlickiARXIV12,KolarPRL12,Gelbwaserumach} in a quantized setup, since nonadiabatic (fast) evolution may yield much higher maximal power than standard stroke cycles that obey the quasiadiabatic Curzon-Ahlborn bound \cite{AlickiJPA79,CurzonAJP75}. The present  scenario has become timely, since the  quantum state-preparation of the mechanical oscillator is now experimentally-feasible by optical pulses  \cite{vannerpnas11,rimbergnjp14}.

\textbf{Model}
We start from  the basic optomechanical Hamiltonian wherein an optical cavity  mode
(denoted by O), is coupled  to (cold and hot) two thermal baths and to  a mechanical oscillator (denoted by  M)

\begin{figure}
	\centering
		\includegraphics[width=1\textwidth]{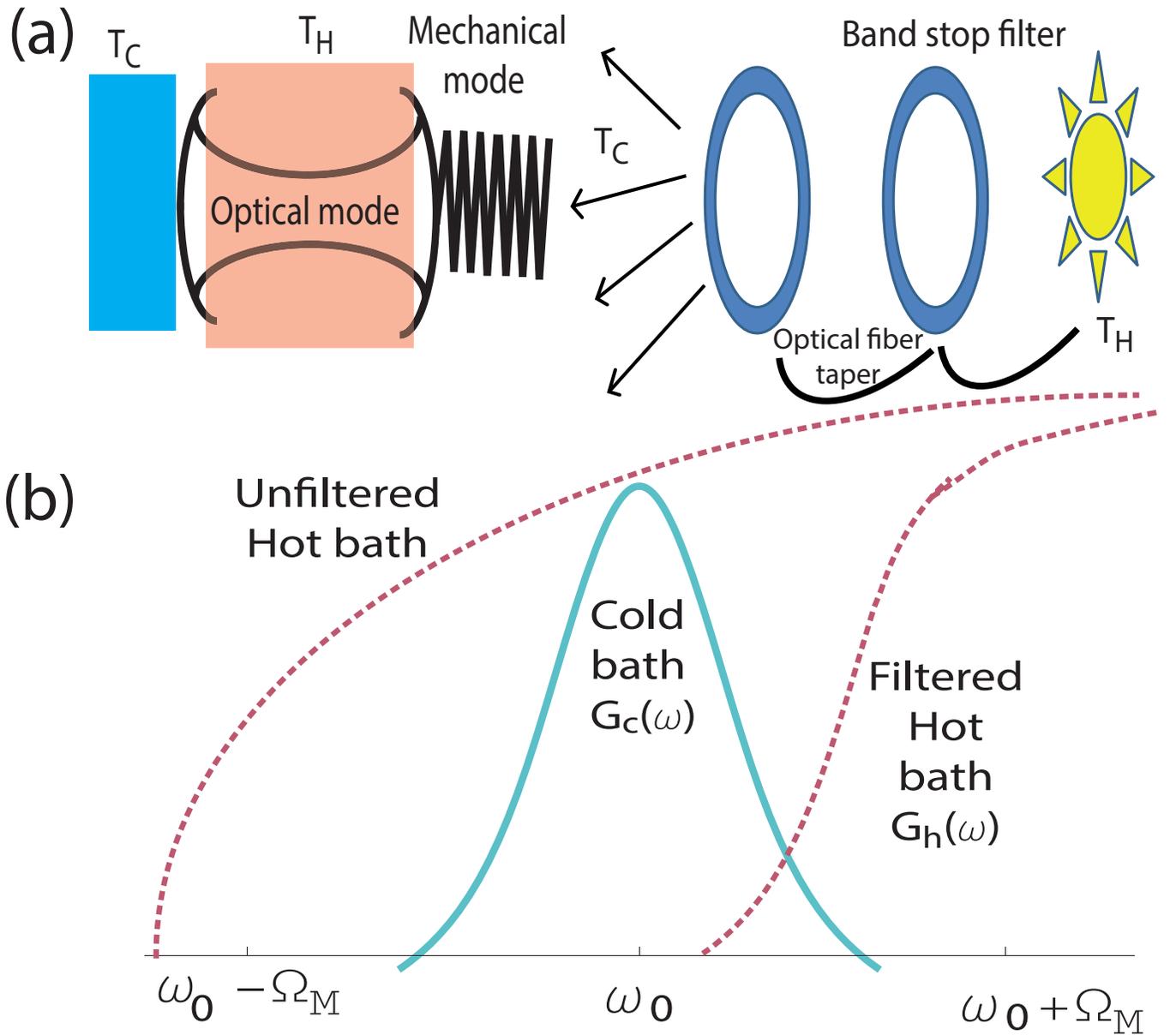}
	\caption{(a) Schematic composition of an autonomous optomechanical implementation of a quantum heat engine: the optical cavity mode (working fluid) is permanently coupled to hot (H) and cold (C) baths and to the mechanical oscillator.  (b) Schematic description of a ring-cavity optomechanical setup coupled to a hot bath ( spectrally filtered by a stop-band filter) and a cold bath (whose coupling spectrum is determined by the cavity  linewidth).}
	\label{fig:OHE}
\end{figure}

\begin{gather}
H_{Tot}=H_{O+M}+(O^{\dagger}+O)\otimes(B_{c}+B_{h}) \notag;\\
H_{O+M}=\omega_{O}O^{\dagger}O+\Omega_{M}M^{\dagger}M+gO^{\dagger}O(M+M^{\dagger}). 
\label{eq:H}
\end{gather}

Here $O^{\dagger},O$ and $M^{\dagger},M$  are the creation and annihilation operators of
the cavity mode and the oscillator, respectively; $\omega_O$, $\Omega_M$ and $g$ are their respective frequencies and coupling rate. The cavity mode O directly interacts with two thermal baths: hot ($B_h$) and cold ($B_c$) baths with non-flat spectra, whereas the oscillator M is damped by a nearly-Markovian phonon bath.
   The response (coupling) spectra of the cold (c) and hot (h) baths may be \textit{broad}, as opposed to a single-mode laser, but controllable, as detailed below:  we shall take the hot bath to be a \textit{spectrally-filtered broadband} source of thermal noise and the cold bath to be the vacuum that is coupled to the cavity mode, with typically a Lorentzian-shaped spectrum \cite{aspelemeyerarxiv13,lorchprx14} (Fig \ref{fig:OHE}b).
	
	In particular,  a toroid microcavity is a candidate for implementing the model mentioned above.  While the microcavity losses are described by the coupling to the cold bath, a second hot bath could be coupled to the cavity through an optical fiber taper (see fig \ref{fig:OHE}(a)).

\textbf{Analysis}
We transform the operators to those of  $\widetilde{O}$, $\widetilde{M}$, the mixed optical-mechanical modes that diagonalize $H_{O+M}$ without changing their frequency (see Methods). Then  the interaction between the optical mode  and the baths in \eqref{eq:H} is found to \textit{indirectly} affect  the mechanical oscillator, enabling it to draw energy from the heat bath via the optical mode. 
Whereas $\widetilde{O}$ rapidly reaches a steady state, $\widetilde{M}$ keeps evolving thereby allowing its amplification.  The evolution equation of their joint state has the form  \cite{AlickiARXIV12,KolarPRL12,Gelbwaserumach}

\begin{equation}
\frac{d\rho_{\widetilde{O}+\widetilde{M}}(t)}{dt}=\sum_{q=0,\pm1}(\mathcal{L}_{q,h} +\mathcal{L}_{q,c})\rho_{\widetilde{O}+\widetilde{M}}(t).
\label{ME_gen}
\end{equation}
 \noindent Here  $q=0,\pm 1$ label the harmonics  $\omega_O, \omega_{\pm}=\omega_O\pm\Omega_M$, respectively, 
and the Lindblad generators  associated with these harmonics in  the two baths, $\mathcal{L}_{q}^{j}$ ($j=h,c)$, depend on the bath-response rates  $G_j(\omega_q)$ (see Methods)
In what follows we shall restrict ourselves to low excitations and \textit{linear amplification} of $\widetilde{M}$ and to  the weak  optomechanical  coupling  regime.
Namely, we shall assume 
\begin{gather}
\left(\frac{g}{\Omega_{M}}\right)^2 \langle n_{\widetilde{M}}\rangle<<1, \quad \frac{g^{2}}{\Omega_{M}}\langle n_{\widetilde{O}}\rangle^2t<<1,\label{regime}
\end{gather}

\noindent where $\langle n_{\widetilde{M}}\rangle$ and $\langle n_{\widetilde{O}}\rangle$ are the mean numbers of quanta in $\widetilde{M}$ and $\widetilde{O}$, respectively.
In this (quasi steady-state, linear-amplification) regime  we can write a  master equation for
the slow dynamics of  $\widetilde{M}$ (see Methods). 
Upon representing the reduced density matrix of $\widetilde{M}$, $\rho_{\widetilde{M}}=Tr_{\widetilde{O}}\rho_{\widetilde{M}+\widetilde{O}}$ in terms of coherent
states $\{|\beta\rangle;\beta\in\mathbf{C}\},$ 
$
\rho_{\widetilde{M}}=\frac{1}{2\pi}\int_{\mathbf{C}}d^{2}\beta\, \mathbf{P}(\beta)|\beta\rangle\langle\beta|,\label{Prep}
$
 where $\mathbf{P}(\beta)$ is the quasi-probability distribution, this  linearized master
equation assumes the form of the Fokker-Planck equation 

\begin{subequations}
\begin{gather}
\frac{d\mathbf{P}}{dt}=\frac{\gamma+\Gamma_M}{2}(\frac{\partial}{\partial\beta}\beta+\frac{\partial}{\partial\beta*}\beta*)\mathbf{P}+d\frac{\partial^{2}\mathbf{P}}{\partial\beta\partial\beta*};\\
\gamma=\frac{g^{2}}{\Omega_{M}^{2}}\Big(G(\omega_{+})\langle n\rangle_{\widetilde{O}}+G(-\omega_{-})\langle n+1\rangle_{\widetilde{O}}-
\left(G(\omega_{-})\langle n\rangle_{\widetilde{O}}+G(-\omega_{+})\langle n+1\rangle_{\widetilde{O}}\right)\Big);\\
d=\frac{g^{2}}{\Omega_{M}^{2}}\left(G(\omega_{-})\langle n\rangle_{\widetilde{O}}+G(-\omega_{+})\langle n+1\rangle_{\widetilde{O}}\right)+d_M.\label{M3}
\end{gather}
\label{eq:fp}
\end{subequations}

\noindent Here $G(\omega)=G_c(\omega)+G_h(\omega)$ and $\Gamma_M$ and $d_M$ are the drift and diffusion rates produce by the \textit{direct} interaction between $\widetilde{M}$ and a phonon bath, while   $\gamma$ and $d$, are their counterparts due to the \textit{indirect} coupling between M and the hot and cold bath, through O. They depend on the combined spectral response (coupling spectra) of the cold and hot baths, $G(\omega)=G_c(\omega)+G_h(\omega)$, sampled at the combination frequencies of O and M, $\pm\omega_{\pm}=\pm(\omega_0\pm\Omega_M):$, and on the mean-number of quanta, $\langle n_{\widetilde{O}} \rangle$  at steady-state.  While $\Gamma_M$ is always positive, $\gamma$ may be also negative: 
  $\gamma$ is  a sum  of  terms that involve the joint response  of the two baths associated with the  $q=\pm 1$  harmonics. Spectral separation of the two baths allows the negativity of the sum, as required for work extraction.
 When there is only one thermal bath with inverse temperature $\beta$ and spectrum $G(\omega)$, $\gamma$ is positive definite, i.e. no work is allowed.

\textbf{Energy amplification}
The energy evolution of any initial state of $\widetilde{M}$ is found from  \eqref{eq:fp} to be

\begin{gather}
\langle E(t) \rangle_{\widetilde{M}}= 
\Omega_{M}(\frac{d(1-e^{-(\gamma+\Gamma_M) t})}{|\gamma|+\Gamma_M}+e^{-(\gamma+\Gamma_M) t}\langle n(0)\rangle{}_{\widetilde{M}})
\label{eq:Em}
\end{gather}

\noindent where $\langle n(0)\rangle_{\widetilde{M}}$ is the mean  initial number of oscillator quanta. Typically  
$
|\gamma|\simeq\frac{g^{2}}{\Omega_{M}^{2}}\bar{G}(\langle n_{\widetilde{O}} \rangle+1)
$ where $\bar{G}$ is the  averaged  response bandwidth of the two baths coupled to the optical mode.

The condition $\gamma+\Gamma_M<0$, where $\Gamma_M$ is the oscillator damping rate by its environment (and not by the optical mode \cite{aspelemeyerarxiv13,lorchprx14}),  ensures the amplification of the oscillator energy,
 $<E(t)>_{\widetilde{M}}$. However, it \textit{does not} represent work extraction: $\langle E(t) \rangle_{\widetilde{M}}$ is ``blind'' to the quantum state of the oscillator and does not discern work from heat (or noise) amplification.In what follows we monitor work extraction  by the  quantized oscillator and analyze its dependence on the evolving quantum state,  based on nonpassivity. 

\textbf{Work  capacity and extraction as nonpassivity}
For a given state of the oscillator $\rho_{\widetilde{M}}$, \textit{the work capacity is the maximum  extractable work} expressed by

\begin{equation}
W_{Max} (\rho_{\widetilde{M}})= \langle E_{\widetilde{M}}(\rho_{\widetilde{M}}) \rangle - \langle (E_{\widetilde{M}}(\rho_{\widetilde{M}}^{pas})\rangle
\label{eq:maxw}
\end{equation}

\noindent Here $\rho_{\widetilde{M}}^{pas}$ is a  \textit{passive state} \cite{LenJSP78,puszcmp1978}, defined as a state
 that minimizes the mean energy of $\widetilde{M}$, without changing its entropy,  and thus \textit{maximizes}  the  work extractable from  the state at hand, $\rho_{\widetilde{M}}$. Equivalently  $W_{Max}(\rho_{\widetilde{M}}^{pas})=0$ , i.e., a passive state  is a state in which work cannot be extracted. The passivity of a state  is manifest by the \textit{monotonic decrease of its energy distribution  $\mathbf{P}(E)$ from its value at the origin, $E_{\widetilde{M}}=0$} \cite{LenJSP78,puszcmp1978}. Nonpassivity  will be shown below to differ from known characteristics of quantum states, such as their purity or Wigner-function negativity \cite{ScullyBOOK97,Schleichbook01}.

As the initial state of the oscillator, $\rho_{\widetilde{M}}(0)$, evolves (via a master equation \cite{lindbladcmp75})   to a  state $\rho_{\widetilde{M}}(t)$, the  maximum extractable work changes, according to Eq. \eqref{eq:maxw},  by 
$\Delta W_{Max}(t)=W_{Max} (\rho_{\widetilde{M}}(t))-W_{Max}(\rho_{\widetilde{M}}(0)).
$
For an increase of the work capacity  with time, it is necessary to prepare M in a \textit{nonpassive}  state   and ensure that $\gamma+\Gamma_M<0$.

The  \textit{upper bound} for $W_{Max} (\rho_{\widetilde{M}})$  is obtained  by taking \textit{the lower bound} of  the second term in Eq. \eqref{eq:maxw}, i.e., setting $\langle E_{\widetilde{M}} (\rho_{\widetilde{M}}^{pas})\rangle  =\langle E_{\widetilde{M}}\rangle _{Gibbs},$
  since  the  Gibbs state 
 is the minimal-energy state with the same entropy as $\rho_{\widetilde{M}}$ \cite{CallenBOOK85,GemmerBOOK10}.
 This passive (\textit{effective} ``Gibbs'') state may be written as 
$
\tilde{\rho}_{\widetilde{M}}^{Gibbs}=Z^{-1}e^{-\frac{\widetilde{H}_M}{T_M}}
$.
Its \textit{effective}  temperature $T_M$ is  merely a parameter that characterizes the evolution of an \textit{arbitrary} $\rho_{\widetilde{M}}(t)$.
	
Upon taking the time derivative of this upper bound of Eq. \eqref{eq:maxw} and using the properties of $\tilde{\rho}_{\widetilde{M}}^{Gibbs}$,  we find that the  \textit{extractable power}  is maximized by

\begin{gather}
\left(\frac{dW}{dt}\right)_{Max}=
\frac{d\langle E_{\widetilde{M}}\rangle}{dt} -T_{\widetilde{M}} \frac{dS_{\widetilde{M}}}{dt}; \quad
 T_{\widetilde{M}}\frac{dS_{\widetilde{M}}}{dt}=\frac{d\langle E_{\widetilde{M}}\rangle_{Gibbs}}{dt},
\label{eq:pnonpas}
\end{gather}

\noindent where $\frac{dS_{\widetilde{M}}}{dt}$ is  the entropy-production rate.

 Equation \eqref{eq:pnonpas}   yields the  efficiency bound upon dividing the output power $\left(\frac{dW}{dt}\right)_{Max}$ by the heat-current input flowing from the hot bath \cite{AlickiARXIV12}, $J_h=\sum_{q=0,\pm1}Tr(\widetilde{H}_O+\widetilde{H}_M)\mathcal{L}_{q,h}\rho_{\widetilde{O}+\widetilde{M}}$ where the sum is over the harmonics $q$, $\mathcal{L}_{qh}$ are the corresponding Lindblad operators associated with $B_h$ and $\rho_{O+M}$ is the joint $O+M$ density matrix. We then obtain the efficiency bound in the form 
\begin{gather}
\eta= 
\frac{\left(\frac{dW}{dt}\right)_{Max}}{J_h}=
\frac{\frac{d\langle E_{\widetilde{M}}\rangle}{dt}-T_{\widetilde{M}}\frac{dS_{\widetilde{M}}}{dt}}{J_h}>0.
\label{eq:pasbound}
\end{gather}

\noindent The  term $-T_{\widetilde{M}} \frac{dS_{\widetilde{M}}}{dt}$ on the r.h.s of \eqref{eq:pasbound}, represents the \textit{effective} heating rate  which cannot be ignored for \textit{a quantum oscillator}: it expresses the rate of loss of  nonpassivity.

\textbf{Work extraction dependence on the quantum state}
By contrast to energy extraction,   \textit{ work-capacity increase} ($\Delta W_{Max}(t)>0$) requires 
 an \textit{initially non-passive distribution} in the  (amplification) regime $\gamma +\Gamma_M<0$. We seek the conditions for maximal work extraction. A clue is provided upon introducing the low-temperature approximation to the entropy production rate in Eqs. \eqref{eq:pnonpas},\eqref{eq:pasbound} for $T_M\approx0$

\begin{equation}
\frac{dS_{\widetilde{M}}}{dt} \approx  (\gamma + \Gamma_M+2d) (\langle \widetilde{M}^\dagger \widetilde{M} \rangle -\langle \widetilde{M}^\dagger  \rangle \langle \widetilde{M} \rangle )+d.
\label{eq:sdotapprox}
\end{equation}

\noindent The first term is strongly state-dependent, as shown in what follows.

\textit{1) }
An initially  \textit{coherent state}, $|\beta(0)\rangle$,  evolves in the linear amplification regime of  the Fokker-Planck equation (\ref{eq:fp}a)  towards a distribution that is centered at an exponentially growing $\beta_{\widetilde{M}}(t)$ and increasingly broadened by diffusion.
The corresponding  maximal  work extraction is given by 
$
W_{\widetilde{M}}=\Omega_M|\beta_{\widetilde{M}}(0)|^{2}e^{-(\gamma +\Gamma_M)t}.
$ 
Thus, the coherent-state work capacity  \textit{exponentially increases}  in this regime. The heating term ($T_{\widetilde{M}}\frac{dS_{\widetilde{M}}}{dt}$) is minimized by this state at short times (according to \eqref{eq:sdotapprox}) and   yields the \textit{optimal condition}  for work extraction. It is sustainable at long times, as an initial coherent state retains its nonpassivity and is  never fully thermalized. 

  For $|\beta_{\widetilde{M}}(0)|\sim 1$ the maximal efficiency bound $\eta$ in Eq. \eqref{eq:pasbound}  may \textit{exceed} the standard Carnot bound, due to   the  slow rising entropy and effective temperature $T_{\widetilde{M}}$. Eventually, the efficiency will drop below Carnot, since the effective temperature of $\rho^{Gibbs}_{\widetilde{M}}$ rises due to diffusion, as
$
1/T_{\widetilde{M}}=\frac{Log(\frac{1+d t}{dt})}{\Omega_M}.
$ 
It is nevertheless significant that an initial small-amplitude coherent state   allows to extract work over many cycles with an efficiency \textit{above the standard two-bath Carnot bound} $1-\frac{T_c}{T_h}$ . The extra efficiency  has its origin in  an extra thermodynamic resource (not present in the standard Carnot engine) that  boosts the efficiency while  complying with the second law of thermodynamics (see SI).
In the quasiclassical  limit  $ |\beta_{\widetilde{M}}(0)| \gg 1$  the Carnot bound is recovered.
Namely, the maximal   power extraction determined by nonpassivity reproduces in the quasiclassical limit that of an \textit{externally} (parametrically) modulated heat engine (proposed in \cite{Gelbwaserumach}) that obeys the standard cyclic work definition \cite{EpsteinNAT95}.

2) The evolution of an \textit{initial phase-averaged coherent state} is obtained by integrating over the initial phase $\theta$ of a coherent state.
 $|\beta_{0}\rangle=|\beta_{\widetilde{M}}(0)|e^{i\theta},$ yielding 
$
\mathbf{P}(\beta',\beta^{*'},t||\beta_0|,0)\propto e^{-|\beta'|^2}(1+|\beta'_0|^2|\beta'|^2)
$
where $|\beta'_0|^2=\frac{4|\beta_0|^2}{d t}$. For   $|\beta'_0|^2>1$ this distribution is \textit{nonpassive}, allowing exponentially growing work extraction.

3) The Fock- state  initial  work capacity ($\Omega_M n_M(0)$)  does not increase with time, but rather decreases until the state becomes passive. The reason is the \textit{fast thermalization} of a \textit{Fock state}:, its heating rate prevails over the rate of   work production, so that the overall change in work capacity  by an initial Fock state is always \textit{negative} (Fig. \ref{fig:workvsself}).

Thus, as opposed to energy amplification (Eq. \eqref{eq:Em}),  the extractable work crucially depends
on the initial  phase-plane distribution  of the piston.
 Other nonpassive distributions, such as squeezed states or Schroedinger-cat states, can be shown to undergo faster entropy production, and are therefore less optimal as far as work production is concerned.

\textbf{Comparison with self-induced oscillations} The regime of work extraction in an autonomous optomechanical heat engine (OHE) differs from the regime  of self-induced oscillations (SIO) in its laser-powered counterpart in several salient respects:

\begin{figure}
	\centering
		\includegraphics[width=1\textwidth]{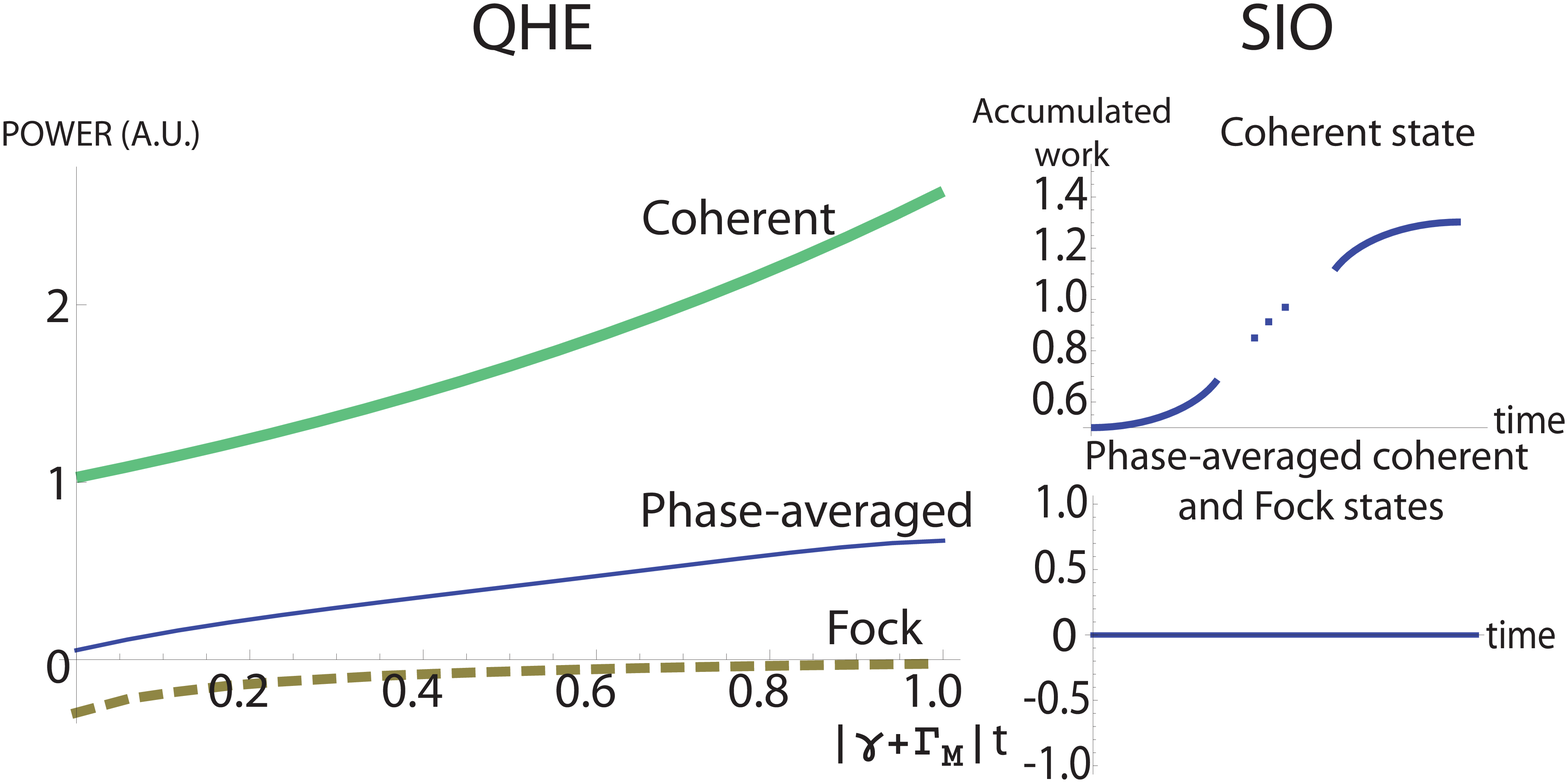}
	\caption{Left: Work extraction dependence on the initial state of the oscillator in the proposed OHE. Right: same in the SIO regime, which relies on phase-locking of the initial position $x(0)$ and the driving amplitude.}
	\label{fig:workvsself}
\end{figure}

1)The OHE regime requires the coupling of the cavity mode to two distinct heat baths to ensure quasi-cyclic operation ($\widetilde{O}$ evolves cyclically  while $\widetilde{M}$ does not),  with clear \textit{thermodynamic} bounds. The SIO-regime does not require such considerations.

2) The SIO regime relies on a single-mode laser drive and does not occur under broadband driving. By contrast, OHE may be powered by  broadband (albeit spectrally-filtered) hot and cold baths (Fig. \ref{fig:OHE}).

3) Perhaps most importantly, the SIO regime involves coherent (phase-locked) oscillator- and cavity- modes, whereas work extraction in the OHE may occur for a \textit{phase-averaged} coherent-state of the oscillator, i.e., without any phase locking with another mode. 
Namely, work extraction, just as energy-extraction or SIO, requires $\gamma+\Gamma_{M}<0$.
Yet, contrary to SIO, even in the classical limit  the mechanical position need not  evolve as $x=x(0) +ACos\Omega_Mt$, but can follow more general trajectories, such as those corresponding to a phase-averaged ensemble. On the other hand, the  restriction of work extraction to nonpassive states has no counterpart in the SIO regime.

\textbf{Discussion}
In the present paper  we have proposed an autonomous (self-contained) optomechanical    heat-engine, allowing for  the  hitherto unexplored role of the state of the quantized mechanical oscillator in the linear amplification regime, i. e., before the onset of saturation for large oscillator amplitudes.

These predictions may be tested for an optomechanical setup that  may be  powered by thermal noise  filtered to eliminate its spectral overlap with the optical cavity mode. Desirable parameters are $(\frac{g}{\Omega_M})^2(\langle n_{\widetilde{O}} \rangle +1)  \bar{G}(\omega_0)\simeq|\gamma|>\Gamma_M.$  Taking $(\frac{g}{\Omega_M})^2(\langle n_{\widetilde{O}} \rangle +1)\sim 0.1$, the requirement amounts to $\bar{G}(\omega_O)\gtrsim 10 \Gamma_M.$ As an example, we take a mechanical oscillator with damping $\Gamma_M\sim$ 1 kHz, frequency $\Omega_M\sim$ 3 MHz, optomechanical coupling $g\sim 1 MHz$, $\langle n_{\widetilde{O}} \rangle \ll 1 $ (Boltzmann factor or mean thermal occupancy) of the optical cavity mode. The filtered heat  bath is coupled to the cavity via band stop filter that can have an unlimited bandwidth but a sharp lower cut off of few MHz width Fig. \ref{fig:OHE}(b).  The EM vacuum has a cavity bandwidth $\sim 1$ GHz.
\textit{Work-extraction rate has been} related to the nonpassivity of the oscillator state, the only rigorously justifiable measure of work  extraction in \textit{time-independent} autonomous setups \cite{LenJSP78,puszcmp1978}. It is shown here to crucially   depend on the initial quantum state,\textit{ in contrast to mean-energy amplification. }
 The resulting efficiency bound \eqref{eq:pasbound} involves  the  \textit{effective} temperature $T_{\widetilde{M}}(t)$. As long as $T_{\widetilde{M}}(t) <T_c$,  Eq. \eqref{eq:pasbound}   may surpass the \textit{standard  two-bath }Carnot bound $1-\frac{T_c}{T_h}$.  Because it complies with Spohn's inequality \cite{SpohnJMP78} (see SI),  the present (evolving) efficiency bound is  consistent with the second law. It shows that  the piston may  serve as a \textit{low-entropy resource} which is excluded by  the standard (classical-parametric) limit of work extraction.   
It comes about  upon allowing for the \textit{inevitable} but commonly ignored  entropy growth of the quantum oscillator  and its linear amplification at finite times.

The  efficiency  and work-production rate (power) derived by us 
are both practically and conceptually interesting,  since  the initial \textit{``charging'' of the oscillator } by quantum state-preparation is sought to be maximally efficient for subsequent operation. 
Such preparation is a one-time investment  of energy   and \textit{does not invalidate} the work and its  extra efficiency obtained  in the linear-amplification regime.
In this respect, our analysis has yielded nontrivial results: (a) \textit{As the initial coherent amplitude of the oscillator decreases, the resulting efficiency increases}, although the \textit{ entropy growth} of the oscillator might then be expected to reduce (rather than enhance) the efficiency. (b) Work extraction obtained from an initial coherent-state has been found to be superior to that of   other states, because of its larger sustainable nonpassivity, conditioned on its low heating or entropy-production rate: This is consistent with  the coherent state being the ``pointer-state'' of the evolution
 \cite{zurekPRD81}.
	(c) Not less remarkable is that, in contrast to laser-powered self-induced oscillations, broadband (heat-) powered work extraction \textit{does not require} coherence or phase-locking: \textit{an initial phase-averaged} coherent state still yields work extraction.

\textbf{Methods}

\textit{The dressing information}

 The dressing transformation can be expressed in terms of new
variables 
\begin{equation}
U=U^{\dagger}UU=e^{\frac{g}{\Omega_M}U^{\dagger}(M^{+}-M)O^{\dagger}OU}=e^{\frac{g}{\Omega_M}(\tilde{M}^{+}-\tilde{M})\tilde{O}^{\dagger}\tilde{O}}.\label{dress1}
\end{equation}
 The operator $O^{\dagger}+O$ which appears in the interaction Hamiltonian
is given in terms of new dynamical variables as 
\begin{equation}
O^{\dagger}+O=\widetilde{O}^{+}e^{\frac{g}{\Omega_M}(\tilde{M}^{+}-\tilde{M})}+e^{-\frac{g}{\Omega_M}(\tilde{M}^{+}-\tilde{M})}\label{sigma_dress}
\end{equation}

The Heisenberg-picture Fourier decomposition of $O^{\dagger}+O$ within
the lowest order approximation with respect to a small parameter $g/\Omega_M$,
can be obtained from 
\begin{gather}
O^{\dagger}(t)=e^{i{H}t}O^{\dagger}e^{-i{H}t}=e^{-i(g\tilde{O}^{\dagger}\tilde{O})^{2}\frac{1}{\Omega_M}t}e^{i\omega_{O}t}\widetilde{O}^{+}e^{\frac{g}{\Omega_M}(\tilde{M}^{+}e^{i\Omega_Mt}-\tilde{M}e^{-i\Omega_Mt})}e^{i(g\tilde{O}^{\dagger}\tilde{O})^{2}\frac{1}{\Omega_M}t}\nonumber \\
\approx e^{-i(g\tilde{O}^{\dagger}\tilde{O})^{2}\frac{1}{\Omega_M}t}\left(\widetilde{O}^{+}e^{i\omega_{O}t}+\frac{g}{\Omega_M}\bigl(\tilde{O}^{\dagger}\tilde{M}^{\dagger}e^{i(\omega_{O}+\Omega_M)t}-O^{\dagger}\tilde{M}^{\dagger}e^{i(\omega_{O}-\Omega_M)t}\bigr)\right)e^{i(g\tilde{O}^{\dagger}\tilde{O})^{2}\frac{1}{\Omega_M}t}\label{eq:fourier-1}
\end{gather}
The approximation made in \eqref{eq:fourier-1} is valid under Eq. \eqref{regime}.

We then have  in the interaction picture

\begin{gather}
H_{O+M}=\widetilde{H}_O+\widetilde{H}_M; \quad \widetilde{H}_O= \omega_{O}\widetilde{O}^{\dagger}\widetilde{O}-(g\widetilde{O}^{\dagger}\widetilde{O})^{2}\frac{1}{\Omega_{M}}; \quad
 \widetilde{H}_M=\Omega_{M}\widetilde{M}^{\dagger}\widetilde{M \notag }\\
O^{\dagger}(t)\approx
\widetilde{O}^{\dagger}(e^{i\omega_{O}t}+\frac{g}{\Omega_{M}}\bigl(\tilde{M}^{\dagger}e^{i(\omega_{O}+\Omega_{M})t}-\tilde{M}e^{i(\omega_{O}-\Omega_{M})t}\bigr)) \notag \\
\widetilde{M}=M-\frac{g}{\Omega_{M}}O^{\dagger}O, \quad \widetilde{O} =O e^{-\frac{g}{\Omega_{M}}(M^{\dagger}-M)O^{\dagger}O}. 
\label{fourier-1}
\end{gather}

\textit{The Master equation}

The master equation \eqref{regime} has the form \cite{AlickiARXIV12,KolarPRL12,Gelbwaserumach}

\begin{gather}
\mathcal{L}_{0,j}\rho_{\widetilde{O}+\widetilde{M}}=\frac{1}{2}\Bigl\{ G_{j}(\omega_{O})\bigl([\widetilde{O}_{-}\rho_{\widetilde{O}+\widetilde{M}},\widetilde{O}_{+}]+
[\widetilde{O}_{-},\rho_{\widetilde{O}+\widetilde{M}}\widetilde{O}_{+}]\bigr)+
G_{j}(-{\omega}_{O})\bigl([\widetilde{O}_{+}\rho_{\widetilde{O}+\widetilde{M}},\widetilde{O}_{-}]+[\widetilde{O}_{+},\rho_{\widetilde{O}+\widetilde{M}}\widetilde{O}_{-}])\Bigr\},\label{generator_loc}
\end{gather}
 
\begin{gather}
\mathcal{L}_{q,j}\rho_{\widetilde{O}+\widetilde{M}}=\frac{g^{2}}{2\Omega_M^{2}}\Bigl\{ G_{j}(\omega_{q})\bigl([W_{q}\rho_{\widetilde{O}+\widetilde{M}},W_{q}^{\dagger}]+[W_{q},\rho_{\widetilde{O}+\widetilde{M}} W_{q}^{\dagger}]\bigr)+
G_{j}(-{\omega}_{q})\bigl([W_{q}^{\dagger}\rho_{\widetilde{O}+\widetilde{M}},W_{q}]+[W_{q}^{\dagger},\rho_{\widetilde{O}+\widetilde{M}} W_{q}]\bigr)\Bigr\}\ ,\ q=\pm1.\label{generator_loc1}
\end{gather}

Here $W_1^{\dagger}= \widetilde{O}^{+}\tilde{M}^{\dagger}$ and $W_{-1}^{\dagger}= \widetilde{O}^{+}\tilde{M}$

The  generator $\mathcal{L}_{0,j}$ drives the evolution of $\widetilde{O}$, which is faster than that of $\widetilde{M}$ whose evolution is  governed by the Lindblad generators $\mathcal{L}_{\pm 1,j}$ .

\textit{The partially stationary regime}

The evolution governed by \eqref{generator_loc} and \eqref{generator_loc1}  has two timescales. The slow one, that includes
all the terms multiplied by $^{\left(\frac{g}{\Omega_M}\right)^{2}}$
is related to changes in the  state of M, while the fast one governs
changes in the system. 

The fast evolution equation for the diagonal elements of the system
is

\[
\dot{\rho}^{nn}_{\widetilde{O}}=-\left((n+1)G(-\omega_{O})+nG(\omega_{O})\right)\rho^{nn}_{\widetilde{O}}+(n+1)G(\omega_{O})\rho^{n+1n+1}_{\widetilde{O}}+nG(-\omega_{O})\rho^{n-1n-1}_{\widetilde{O}}
\]

It reaches quickly  steady state 

\begin{equation}
\tilde{\rho}^{nn}_{\widetilde{O}}=\left(\frac{G(-\omega_{O})}{G(\omega_{O})}\right)^{n}(1-\frac{G(-\omega_{O})}{G(\omega_{O})})
\label{eq:rhonn}
\end{equation}

with average population  $\langle n_{\widetilde{O}}\rangle=\frac{G(-\omega_{O})}{G(-\omega_{O})+G(\omega_{O})}$,
where $G(\omega)=G_{h}(\omega)+G_{c}(\omega)$,
being  the bath response spectra .  Under these conditions, the master equation for $\rho_{\widetilde{M}}=Tr_{\widetilde{O}}\rho_{\widetilde{O}+\widetilde{M}}$ may be rewritten in the Fokker-Planck form  \eqref{eq:fp}.

\section*{Supplementary information}

\renewcommand{\theequation}{S\arabic{equation}}

\section{Negativity of $\gamma$}

The complete expression for $\gamma$

\begin{gather}
\gamma=\frac{g^{2}}{\Omega_{m}^{2}} \langle n+1\rangle_{\widetilde{O}} \times \notag \\
\left[G_{h}(\omega_{+})G_{h}(\omega_{0})\left(e^{-\beta_{h}\omega_{0}}-e^{-\beta_{h}\omega_{+}}\right)+G_{c}(\omega_{-})G_{c}(\omega_{0})\left(e^{-\beta_{c}\omega_{-}}-e^{-\beta_{c}\omega_{0}}\right)+G_{h}(\omega_{-})G_{h}(\omega_{0})\left(e^{-\beta_{h}\omega_{-}}-e^{-\beta_{h}\omega_{0}}\right)\right.+ \notag\\
G_{c}(\omega_{+})G_{c}(\omega_{0})\left(e^{-\beta_{c}\omega_{0}}-e^{-\beta_{c}\omega_{+}}\right)+G_{c}(\omega_{+})G_{h}(\omega_{0})\left(e^{-\beta_{h}\omega_{0}}-e^{-\beta_{c}\omega_{+}}\right)+G_{h}(\omega_{-})G_{c}(\omega_{0})\left(e^{-\beta_{h}\omega_{-}}-e^{-\beta_{c}\omega_{0}}\right) \notag\\
\left.+G_{h}(\omega_{+})G_{c}(\omega_{0})\left(e^{-\beta_{c}\omega_{0}}-e^{-\beta_{h}\omega_{+}}\right)+G_{c}(\omega_{-})G_{h}(\omega_{0})\left(e^{-\beta_{c}\omega_{-}}-e^{-\beta_{h}\omega_{0}}\right)\right]
\end{gather}

where only the terms in the last line may be negative. 

Physically, they represent processes where the $\widetilde{O}$ mode  interacts with high frequency modes of the hot bath and low frequency modes of the cold bath. 
Under the right combination of frequencies and temperatures, a high-frequency excitation is absorbed from the hot bath and a low-frequency excitation is emitted to the cold bath. The energy differences is received by the mechanical mode $\widetilde{M}$, partly is dissipated as heat (increasing the entropy) and partly is extracted as work.

  To ensure  the  negativity of $\gamma$, the spectra should be engineered to increase the relative weight of the last-line terms compared to the other terms.

\subsection{Passivity and non-passivity of the phase-plane distribution}
First, assume \textit{an initial passive distribution}, i.e., an \textit{isotropic} distribution satisfying \textit{ monotonic decrease} with energy: $\frac{\partial \mathbf{P}(r_{0})}{\partial r_{0}}<0$. Then, in the $\gamma+\Gamma_M <0$ regime
we find that
$\frac{\partial \mathbf{P}(re^{i\theta},t)}{\partial r}$ is \textit{negative} at any
time, so that $ \mathbf{P}(re^{i\theta},t)$ remains  passive even in the gain regime, always prohibiting \textit{work extraction}.  Hence, \textit{state-passivity is preserved} by the Fokker-Planck phase-plane evolution. 

A notable example of passive-state evolution is that of an initial  thermal state, whose evolution is given by 

\begin{equation}
\mathbf{P}(\beta,t|\beta(0),0)=\frac{\gamma}{\pi\left(d(1-e^{-\gamma t})+\sigma\right)}e^{-\frac{\gamma|\beta|^{2}}{d(1-e^{-\gamma t})+\sigma}}
\label{eq:ppas}
\end{equation}

where $\sigma$ is the initial width of the distribution. This state remains thermal (and passive) at any time. 
Although no work is extracted, the mean  energy of \textit{the thermal state
increases for negative $\gamma+\Gamma_M$}. This  example  clearly shows the difference between energy gain and work extraction (Fig, 1c, Fig. 2a). 

 A unitary operation that transforms this nonpassive  distribution to a Gibbs state, thereby maximizing the work  extraction,   is a displacement of the exponentially growing $\beta$ towards the origin, by $\beta(0)e^{-\frac{\gamma}{2}t}e^{-i\nu t}$,  thereby attaining the transformed distribution 

\begin{gather}
\mathbf{\tilde{P}}(\beta,t|\beta(0),0) \rightarrow\mathbf{\tilde{P}}^{pas}(\beta,t|\beta(0),0)=
\frac{\gamma}{\pi d(1-e^{-\gamma t})}e^{-\frac{\gamma|\beta|^{2}}{d(1-e^{-\gamma t})}}.
\label{eq:transp}
\end{gather}

\subsection{The nonpassivity bound and the Second Law }

Under weak system-bath coupling, $\langle H_{\widetilde{M}} (t) \rangle$ undergoes quasi-cyclic, slowly-drifting evolution which is the nonadiabatic counterpart of Carnot cycles. The steady-state of $\widetilde{O}$ and the slow-changing cycles of $\widetilde{M}$correspond to Markovian evolution of $\widetilde{M}+\widetilde{O}$ \cite{Gelbwaserumach,LindbladBOOK83,SzczygielskyPRE13}.

  The  \textit{bound}  for the total entropy-production rate of $\widetilde{M}+\widetilde{O}$  is provided by the Clausius   version of the second law  in the form of Spohn's inequality that holds  under Markovian evolution \cite{SpohnJMP78}. 
	Assuming a  small ratio of the system-piston coupling strength $g$ to the piston oscillation-energy (frequency) $\Omega_M$, the system and the piston are nearly in a product state,  their   production of entropy is even closer to being additive: 
	
	\begin{equation}
	\rho_{\widetilde{O}+\widetilde{M}}=\rho_{\widetilde{O}}\otimes\rho_{\widetilde{M}}+O(\frac{g}{\Omega_M})^{2}; \quad
	\dot{S}_{O+M}=\dot{S}_{\widetilde{O}}+\dot{S}_{\widetilde{M}}+O(\frac{g}{\Omega_M})^{4}.
	\label{eq:rhosp}
	\end{equation}
	
	Then, considering that after cross-graining,  $\dot{S}_{\widetilde{O}}=0$ at periodic steady-state and the only entropy production is that of the piston, $\dot{S}_{\widetilde{M}}$, the second law expressed by the Spohn inequality reads 

\begin{equation}
\dot{S}_{\widetilde{M}} \geq\frac{J_{h}}{T_{h}}+\frac{J_{c}}{T_{c}}.
\label{eq:2law}
\end{equation}

\noindent In what follows, this inequality will be used to infer efficiency bounds that allow for entropy and work production by $\widetilde{M}$.

\begin{gather}
\eta^{Max}= 
\frac{\left(\frac{dW}{dt}\right)_{Max}}{J_h}=
\frac{\frac{d\langle H_{\widetilde{M}}\rangle}{dt}-T_{\widetilde{M}} \dot{S}_{\widetilde{M}}}{J_h}>0.
\label{eq:pasbound1}
\end{gather}

\noindent The  term $-T_{\widetilde{M}} \mathscr{\dot{S}}_{\widetilde{M}}$ on the r.h.s of \eqref{eq:pasbound1}, reflecting the heating and entropy change of $\widetilde{M}$,  is neglected by  the prevailing semiclassical treatments that treat $\widetilde{M}$ as a classical parametric drive  of $\widetilde{O}$,  but
$ \dot{S}_{\widetilde{M}}$ cannot be ignored for \textit{a quantum piston},  as shown below. Despite its being ``fictitious'' or effective, the product $T_{\widetilde{M}} \dot{S}_{\widetilde{M}}(t)$ is a faithful measure of the piston heating rate, because it expresses the rate of its passivity increase (or nonpassivity loss).

The compliance of \eqref{eq:pasbound1} with the standard Carnot bound is only  ensured if $T_c \leq T_{\widetilde{M}}$. Yet for  $T_{\widetilde{M}} < T_c$,  the Spohn inequality  \eqref{eq:2law} 
 implies   that  Eq.  \eqref{eq:pasbound1} satisfies

\begin{gather}
\eta^{Max} (T_{\widetilde{M}} < T_c)\leq 1-\frac{T_{\widetilde{M}}}{T_{h}}.
\label{eq:tpsmall} 
\end{gather}
 
\noindent  The efficiency in   Eq. \eqref{eq:tpsmall} \textit{surpasses} the standard two-bath Carnot bound, $1-\frac{T_c}{T_h}$, when $T_{\widetilde{M}}<T_c$. Nonetheless, Eq. \eqref{eq:tpsmall} adheres to  Spohn's inequality \cite{SpohnJMP78} and therefore to \textit{the second law}.

\textbf{Acknowledgments}

 The support of ISF, BSF, AERI and CONACYT is acknowledged.

\end{document}